# Translating declarative control elements to imperative using 'l-value redefinition graphs'

Anthony Savidis[1,2]
[1] Department of Computer Science, University of Crete
[2] Institute of Computer Science, FORTH
as@ics.forth.gr

**Abstract.** We focus on control constructs that allow programmers define actions to be performed when respective conditions are met without requiring the explicit evaluation and testing of conditions as part of an imperative algorithm. Such elements are commonly referred as declarative, not theoretically related to declarative languages. We introduce declarative constructs in the C++ language, presenting the translation method to standard C++. The innovative feature of our method is the accommodation of l-values involving arbitrary pointer / array expressions and objects, supporting immediate runtime evaluation upon content update even if such l-values bind to variant storage locations at runtime. To accomplish this we define 'l-value redefinition graphs', capturing storage binding dependencies among variables, being the floor-plan of our code generation and runtime management approach.

## 1   Introduction

Currently, there are various declarative extensions over the imperative programming style one of the most popular being event based programming. The latter enables the declarative association of event handlers to event types, effectively relieving programmers from the burden of explicit imperative polling. Event-based programming introduced in the past a programming paradigm shift, as event handlers implied declarative specifications of the form "**on** *event* **do** *stmt*" eliminating the necessity for polling. Although such new programming models are not a panacea, the reduction of control-flow dependencies, putting emphasis on establishing links among actions and preconditions, may turn it to beneficial for specific domains.

For instance, User Interface development involves such declarative associations and dependencies, a fact justified by the early existence of many languages departing from the strict imperative style, introducing elements like constraints [2,3], event response systems [6], and pre-conditional activation of agents [1,7]. The need for declarative control constructs in User Interface languages was identified very early [4], while many popular software design patterns [5], like Decorator, Façade, Factory, View, emerged from research on User Interface development systems.

The work reported in this paper relied on the development of a User Interface programming language [10] compiled to standard C++, offering declarative programming elements, while supporting domain-specific classes and a C-style programming kernel with pointers and dynamic memory allocation. It is believed that the incorporation of declarative elements in the C++ language is beneficial, enabling programmers to practice novel programming patterns, assuming we acknowledge that User Interface programming is not a task merely attributed to domain-specific languages. We will discuss the implementation of three categories of declarative elements for the C++ language, as shown in **Table 1**.

**Table 1.** The categories of declarative constructs introduced in the C++ language.

| Element | Explanation |
|---|---|
| *Unidirectional constraint* **α := β** | <ul><li>**α** is an *l-value*, **β** is an *r-value*</li><li>Introduces a dependency link **α←β$_i$** with every variable **β$_i$** syntactically involved in **β**.</li><li>Multiple constraints for the same *l-value* are stacked according to their order of instantiation, distinguishing the *top* constraint.</li><li>The *top* constraint is forced as follows: *if any variable syntactically involved in **β** is modified, the assignment **α=β** is directly performed.*</li></ul> |
| *Variable monitor* **α :=: stmt** | <ul><li>**α** is an *l-value*, **stmt** is a language statement.</li><li>Multiple monitors for the same *l-value* are stacked according to their order of instantiation, distinguishing the *top* monitor.</li><li>The *top* monitor is called as follows: *after **α** is modified, the **stmt** is directly performed, while the monitor is disabled during its execution.*</li></ul> |
| *Pre-conditional statement* **β ?? stmt** | <ul><li>**β** is an *r-value*, **stmt** is a language statement.</li><li>The precondition is evaluated as follows: *if any variable syntactically involved in **β** is modified, the precondition **β** is evaluated, and if true, the **stmt** is directly performed.*</li></ul> |

Declarative elements are essentially declared objects, automatically applying the logic shown at the right column of **Table 1** upon instantiation. They may be defined either at file scope (global) or in the body of classes (local). In the former case a single instance is created, following the semantics of global initialization, while in the latter case a distinct instance is produced with every *owner* class instantiation. The symbols shown, like **:=** or **:=:**, are only indicative, to avoid the need for extra keywords and syntactic conflicts with the rest of C++ expressions.

Following our definitions, re-evaluation of expressions is performed *only* due to updates of variables syntactically encompassed within such expressions. For instance, given a constraint of the form *x := f()*, *f* implemented as *{ return y }* and *y* a global variable, *f()* will not be re-evaluated when *y* changes. In this case there is always an equivalent more modular wraparounds, by either transforming *f* so that *y* becomes an actual argument, or by alternatively substituting the constraint with a monitor of the form *y :=: x=f()* having the same semantic effect.

Additionally, as we will discuss at the end, constraints like *x=y.f()*, *y* an object with member function *f*, are also handled in our approach, meaning *y.f()* will be re-evaluated when *y* changes, requiring a couple of assumptions to qualify *y* content updates as preserving the class invariant. Similarly, to support constraints involving objects, preconditions as defined Design By Contract™ [9] should be respected. Our solution to the latter is the introduction of pre-conditional constraints of the form *x := y.f() given y.pre_f()*, let *pre_f* be the precondition function of *f*.

Throughout the discussion we focus on the code generation method, rather than on the precise semantic details of the declarative constructs, as different versions of those are currently met in numerous types of languages [1,2,3,7], however, without supporting pointers, array subscript expressions and objects.

### 1.1 Overview

We provide a quick briefing of our implementation method. The most demanding issue is the support of all sorts of *l-values*, including pointer variables of any indirection depth, array indexing expressions, and typical object member access. Our method is actually split in two parts (see **Fig. 1**), the discussion primarily focusing on the first: (a) the syntax-directed compilation method, generating imperative elements (C++ code); and (b) the implementation of an augmented C++ type library on top of which the code generation method is built (more details in [10]).

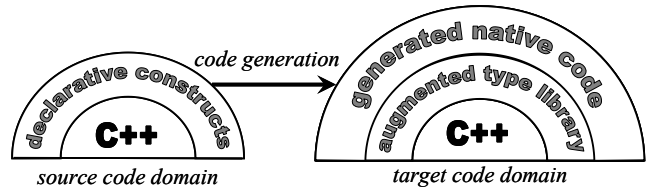

**Fig. 1.** From hybrid C++, with declarative control constructs, to native C++, relying on the augmented type library, and code generation through l-value redefinition graphs.

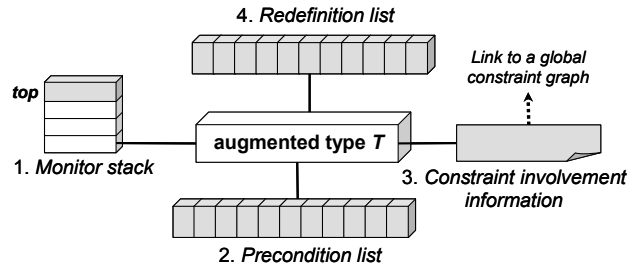

**Fig. 2.** Model (super-class) for augmented types to support involvement in declarative constructs; the lists contain functions which upon call commit their respective action.

The augmented type library consists of template classes, as appropriate adaptations of all built-in and user defined data types, including scalar and aggregate types, so as to support their involvement within declarative constructs. Every augmented type is derived from the basic model illustrated in **Fig. 2**, so that instance variables support runtime bookkeeping of their involvement in constructs like monitors or pre-conditional statements. When such an *augmented variable* is modified, it calls its monitors, it evaluates its pre-conditions and it posts notifications for constraint resolution to a global constraint graph. Also, it performs the necessary actions if value update alters the storage binding of other program variables (i.e. redefinition behavior).

Clearly, from the various lists of functions mentioned in **Fig. 2**, the role of the redefinitions list is less intuitive in comparison to the rest, as we have not yet formally defined the notion of 'redefinition'. Informally, a redefinition is a relationship holding among two l-values if the value of one affects the storage binding of the other. For instance, given two l-values *A[i]* and *\*p*, *i redefines A[i]* and *p redefines \*p*. The importance of this relationship in declarative constructs lays on the fact that redefinition dependent l-values involved in declarative constructs may be mapped to variant storage regions at runtime depending on the value updates of redefining variables. Hence, once redefining variables are modified, the runtime system should propagate the update by correctly rearranging the involvement of the redefinition dependent variables within declarative constructs. As a simple example demonstrating this need, consider the constraint *\*p:=x*, where *p* changes at runtime; clearly, the actual constrained object *\*p* varies over runtime depending on the value of *p*.

## 2  Related work

Constraint-based relationships applied on User Interface elements have been introduced in ThingLab [2,3], a system offering a class-oriented language supporting unidirectional constraints over simple variables. Pre-conditional statements have been introduced in the context of the Serpent User Interface Management System (UIMS) [1], latter commercialized as the Alpha UIMS [7], relying on the assumption that any imperative program doing something meaningful is predicated on the fact that variables change. Effectively, when such changes occur, other things change in response. In this context, dependency-based programming expresses these relationships and cascades of data in a simple and straightforward manner via unidirectional constraints and precondition based activation of domain-specific classes called view controllers. The Slang language of the Alpha UIMS compiled to byte code that was interpreted by a virtual machine, while it supported only trivial variables (no pointers, no array indexing, and no objects).

Algorithms for integrating pointer variables within unidirectional constraint models were introduced in [12]. The technique proposed a centralized reasoning algorithm about indirection dependencies, however, not addressing the issue of l-value redefinition and object variables.

While there are numerous other similar approaches regarding the introduction of declarative elements in domain-specific languages, mostly varying with respect to the specification style and the host language, we are not aware of efforts towards an integrating import of such elements in mainstream languages, like Java, C++ or C#. The latter is arguably due to the inherent implementation complexity, since all known methods rely on: (a) interpretable techniques over custom-made virtual machines; and (b) runtime management algorithms, where everything needed to accommodate declarative elements is orchestrated only as a runtime activity. Finally, since none of the existing techniques addresses thoroughly all types of l-values met in mainstream languages, especially assuming C++ as the subject language, only incomplete and partial solutions may be possible.

## 3    *l-value* redefinition graphs

A redefinition graph states the dependencies among l-values, regarding their runtime mapping to storage regions. While at any given point in time, an l-value expression unambiguously represents a single storage region, it may denote alternative regions during program trace. For instance, the region implied by *p[i]* is dynamically defined by the values of *p* and *i*. Using interchangeably the terms l-value and variable, we say that an *l-value is redefined*, that is a *variable is redefined*, when, while its scope is active, it is mapped to a new storage region. This occurs once the storage region of variables depends on the value of other variables. We formulate such dependencies, presenting the syntax directed construction of l-value redefinition graphs over the basic C++ l-value grammar rules.

### 3.1    Definitions and examples

Let's consider the non-terminal *Lvalue* for common l-values of the C++ language, *x* and *y* be sentences produced from *Lvalue*, *Lvalue*⇒*x* and *Lvalue*⇒*y*. We define a ***redefinition relationship*** $x \rightarrow y$ as follows:

(i) $x$ pure_substring_of $y$
(ii) $\neg \exists w$: *Lvalue*⇒$w \wedge w$ pure_substring_of $y \wedge x$ pure_substring_of $w$

The first property states that $x \rightarrow y$ is not reflexive, as *x* cannot be equal to *y*. The second property states that *x* should belong to the set of longest strings with the first property. The $x \rightarrow y$ relationship results in acyclic graphs, as for circles, nodes *z* of this circle should be pure substring of themselves. Examples of redefinition graphs are provided in **Fig. 3**. At the graph with label 1, it is observed that there are no direct edges from *x*, *y* and *p* towards *q[f(p[x+y])]*, since the former three l-values are substrings of *p[x+y]*, the latter being the longest pure substring of *q[f(p[x+y])]*.

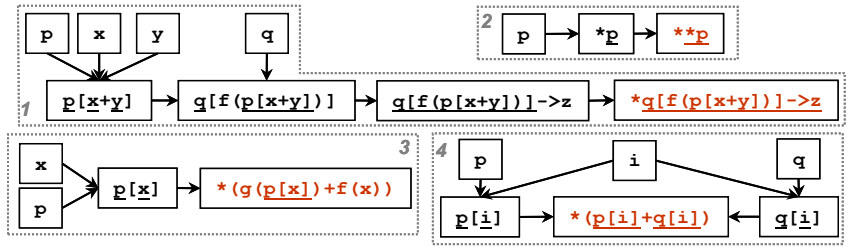

**Fig. 3.** Example redefinition graphs; the longest l-value substrings are shown underlined.

Practically, an edge explicitly connecting *y* to *q[f(p[x+y])]* is unnecessary, since, during runtime, after *y* is modified, the computation of the new value of *p[x+y]* needs to be always carried out first, before computing the new storage of *q[f(p[x+y])]*. Effectively, the redefinition relationship focuses only on the minimally required edges. For the same reason, at the graph with label 3, none of the two occurrences of *x* inside the l-value *\*(g(p[x])+f(x))* leads to an edge directly from *x*, as there is the longer substring *p[x]* already encompassing l-value *x*.

### 3.2 Syntax direct construction

The syntax-directed construction of redefinition graphs is straightforward via simple semantic rules for l-value productions. We define *L* grammar symbol for l-values, with type having a synthesized attribute *redef* as a *list of L*. Semantically, *L.redef* is populated during parsing so that: $\forall\ x \in L.redef : x \rightarrow L$. In other words, the *redef* attribute of an l-value contains all the l-values that redefine it. Similarly, both *E* (expression) and *EL* (expression list) non-terminals have an attribute *lvs* (l-value list) as a *list of L*, corresponding to their encompassing l-values (longest substrings), stored according to their order of appearance. Finally, *L*, *E* and *EL* have a synthesized attribute *str* carrying the input string from which the grammar symbol is actually reduced. The semantic rules are provided in **Table 2**.

**Table 2.** Semantic rules to produce redefintion graphs for most C++ l-values,.

| | |
|---|---|
| $L \rightarrow$ **id** | { *L.str* = **id**; } |
| $L \rightarrow$ *\*E* | { *L.redef* = *E.lvs*; *L.str* = "**\***"+*E.str*; } |
| $L \rightarrow$ *E* -> **id** | { *L. redef* = *E.lvs*; *L.str* = *E.str*+"->"+ **id**; } |
| $L \rightarrow$ *E* **.** **id** | { *L. redef* = *E.lvs*; *L.str* = *E.str* +"**.**" + **id**; } |
| $L \rightarrow E_1$ ->\**$E_2$* | { *L. redef* = *merge*($E_1$.*lvs*, $E_2$.*lvs*); *L.str* = $E_1$.*str* + "->\*" + $E_2$.*str*; } |
| $L \rightarrow E_1$ .\**$E_2$* | { *L. redef* = *merge*($E_1$.*lvs*, $E_2$.*lvs*); *L.str* = $E_1$.*str* + "**.\***" + $E_2$.*str*; } |
| $L \rightarrow E_1[E_2]$ | { *L. redef* = *merge*($E_1$.*lvs*, $E_2$.*lvs*); *L.str* = $E_1$. *str* + "**[**" + $E_2$.*str* + "**]**"; } |
| *Prim* $\rightarrow L$ | { *Prim.lvs* = *list(L)*; *Prim.str* = *L.str*; } |
| $E \rightarrow$ *Prim* | { *E* = *Prim*; } |
| $E \rightarrow E_1$ *op* $E_2$ | { *E.lvs* = *merge*($E_1$.*lvs*, $E_2$.*lvs*); *E.str* = $E_1$.*str*+ *op.str* + $E_2$.*str*; } |
| $E \rightarrow E_1$ (*EL*) | { *E.lvs* = *merge*($E_1$.*lvs*, *EL.lvs*); *E.str* = $E_1$.*Lvs* + "**(**" + *EL.str* + "**)**"; } |
| $EL \rightarrow E$ | { *EL.lvs* = *E.lvs*; *EL.str* = *E.str*; } |
| $EL \rightarrow EL_1$, *E* | { *EL.lvs* = *merge*($EL_1$.*lvs*, *E.lvs*); *EL.str* = $EL_1$.*str*+ "**,**" + *E.str*; } |

The graphs are not constructed as typical graph structures, but are implicitly gained by chaining the *redef* lists of l-values. More specifically, any l-value *x* with a *redef* list containing $y_1, ..., y_k$ represents the sub-graph with edges $x \leftarrow y_1, ..., x \leftarrow y_k$. Together with *redef*, encompassing the redefining l-values, we also need to produce the 'opposite' list of redefinition-dependent l-values - its creation is handled with small extensions on the basic rules and is omitted for clarity. Finally, we need to ensure that for any l-value *x*, *x.redef* contains only the l-values reduced by the longest substrings of *x.str*. Consequently, when merging the *redef* lists for non-terminals on the RHS, we have to drop any l-value that appears to be reduced from a substring (includes equality) of any other l-value within the merged lists. This logic is applied by the *merge* function implemented at the left part of **Fig. 4**.

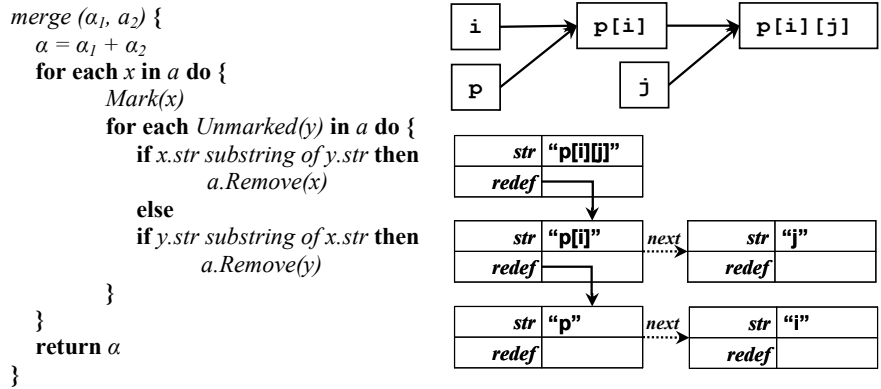

**Fig. 4.** The function to merge two lists of l-values (left), removing l-values reduced from substrings of other l-values of the input lists, and an example (right) of the produced *redef* links for "p[i][j]", assuming 'T** p' pointer variable ('next' links *redef* list elements).

### 3.3 Redefinition graphs and declarative constructs

The actual variables engaged in declarative constructs may vary at runtime when those are redefinition-dependent on variables whose content is changed. If a variable is redefined, two steps need to be taken: (i) exactly before redefinition its involvement in declarative construct has to be cancelled; and (ii) exactly after redefinition, such involvement should be set again. To accommodate the involvement of redefined and redefining variables in declarative constructs, for any $x \rightarrow y$, the modification of $x$ should be handled as follows: (a) *before x is modified*: $y$ cancels its involvement in declarative constructs and requests the same action to all redefinition dependent variables $z$ with $y \rightarrow z$; (b) *after x is modified*: $y$ sets its involvement in declarative constructs and requests the same action to all redefinition dependent variables $z$ with $y \rightarrow z$. This requires the ability to track down during execution content updates of variables, in particular, the point exactly *before* and *after* it is performed, so as to appropriately embed the redefinition-specific actions. An example illustrating such dynamic involvement in declarative constructs is provided in **Fig. 5**, for a constraint $p[i]:=*x$, $p$ an array variable, with redefinition relationships $i \rightarrow p[i]$ and $x \rightarrow *x$.

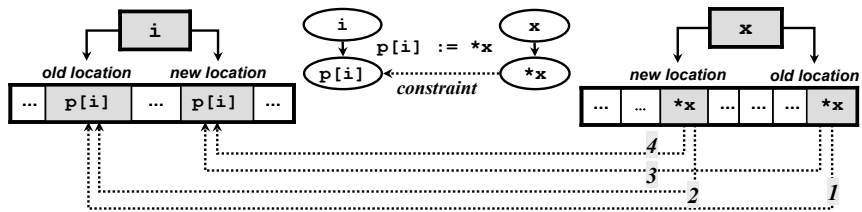

**Fig. 5.** Dynamic reestablishment of constraint links among redefined variables, showing the constraining links (dashed lines) for the each of the four variable update scenarios.

Following **Fig. 5**, initially the constraint labeled '1' holds. Assuming $x$ is firstly modified, *x is redefined, so *x updates it involvement in the constraint to reflect its new storage location, shown by the constraint labeled '2'. Alternatively, if $i$ is firstly modified, *p[i]* is redefined, leading to the constraint with label '3'. Finally, after $i$ and $x$ are modified the active constraint is the one with label '4', no more involving any of the originally involved variables.

In order to support content-update monitoring for program variables, an augmented type library has been designed and implemented, on top of the C++ language, offering all the necessary functionality for update tracking, redefinition behavior, involvement in declarative constructs, and finally the implementation of the semantics of declarative constructs as such (besides constraints that, following the common practice, are managed by a centralized constraint system). We continue by presenting the key features of the augmented type library.

## 4. Augmented type library

### 4.1 Overview

The augmented type library consists of template classes corresponding to scalars, *void\**, arrays, data pointers, and pointers to member or non-member functions. User-defined classes / structures, as discussed at the end, are treated in a special manner. The basic 'augmented' functionality is populated in the context of a super class from which the augmentation templates are derived. We call such augmented types 'smart types', in a way analogous to smart pointers offering a type of 'smart' adaptation over native pointer types.

### 4.2 'initialization', 'action' and 'redefinition' code generated functions

Smart types provide the framework to register, host, and automatically call special-purpose functions that are produced during code generation due to declarative constructs. In **Fig. 6**, the C++ API of the super class from which all template classes are derived is provided, showing the registration API for action and redefinition functions, and the two functions performing the necessary actions exactly before / after content is modified.

```
class SmartTypeSuper { public:
   void ActionsBeforeChange (void);
   void ActionsAfterChange (void);
   void HandleMonitor(MonitorFunc f, bool add, void* owner);
   void ControlMonitors (bool enable);
   void HandlePrecondition(PrecondFunc f, bool add, void* owner);
   void HandleConstraint(AssignFunc f, bool add, void* owner);
   void HandleDependency (SmartTypeSuper* from, bool add, void* owner);
   void HandleRedefinition(RedefFunc f, bool add, void* owner);
};
```

*Redefinition functions, monitors and pre-conditional statement testers are pairs <f, x> of a function address **f** and the variable-owner object **x**.*

*ActionsBeforeChange*() {
   **for each** *redefinition entry* **r do**
      *r.f*(*r.x*, **false**);
}

*ActionsAfterChange*() {
   **for each** *redefinition entry* **r do**
      *r.f*(*r.x*, **true**);
   *ControlMonitors*(**false**); *i.e. disable*
   **for each** *monitor entry* **m do**
      *m.f*(*m.x*);
   *ControlMonitors*(**true**); *i.e. enable*
   **if** *constrains other variables* **then**
      *Call constraint resolution*;
   **for each** *precondition tester* **p do**
      *p.f*(*p.x*);
}

**Fig. 6.** Part of the smart-type super class with the implementation of the *ActionsBeforeChange* and *ActionsAfterChange* key functions.

All functions types of **Fig. 6**, like *MonitorFunc* and *PrecondFunc*, correspond to functions produced due to code generation. We distinguish three key categories of such code generated functions:

- *Initialization functions*. They control (set or cancel) the involvement of variables in declarative constructs. For instance, given a constraint *x:=y*, two such functions need to be generated, one for the involvement of *x* as a constrained variable, and one for the involvement of *y* as a constraining variable.

- *Action functions*. They concern the commitment of declarative constructs when their triggering condition is fulfilled. For example, given the pre-conditional statement *x==y ?? g(x,y)*, a tester function is generated with the block *{ if (x==y) g(x,y) }* to evaluate / perform the pre-conditional statement. The *PrecondFunc*, *AssignFunc* and *MonitorFunc* of **Fig. 6** are action-function types.

- *Redefinition functions*. They encapsulate calls to initialization functions of all redefinition dependent variables. For instance, given a constraint *\*x:=y*, the redefinition function of *x* encompasses appropriate calls to the initialization functions of *\*x*. The registration of redefinition functions for variables like *x* that redefine other variables (like *\*x*), is performed in their corresponding initialization functions. These functions relate to the *RedefFunc* type of **Fig. 6**.

```
template <class T> class SmartScalar : public SmartTypeSuper {
   T val;
   template <class C> void Set (const C& x)
      { ActionsBeforeChange(); val = x; ActionsAfterChange(); }
 public:
   const T Value (void) const { return val; }
   template <class C> const SmartScalar operator=(const C& x)
      { Set(x); return *this; }
   SmartScalar (const SmartScalar& x) : val(x.Value()){}
   template <class C> SmartScalar (const C& x) : val(x){}
};
```

**Fig. 7.** Template class for augmented scalar types.

An example of an augmented type is the *SmartScalar* template class for scalar types of **Fig. 7**, showing the way variable update monitoring is handled during runtime: the content of variables may be changed only through the overloaded assignment operator, the latter calling *Set* which calls *ActionsBeforeChange* before content modification is committed and *ActionsAfterChange* exactly after the content update assignment *val = x* is performed. The content-update 'capture' behavior applies directly on the generated code since for scalar variables involved as modified l-values in assignments, the overloaded assignment operator applies. Additionally, for all r-values, except those used as initializers to reference variables, a call to *Value()* is generated to extract the scalar typed content. Pointer types are more comlpictated, following a technique reported in [11].

**Code generation example** A detailed example follows presenting the way code generation is orchestrated so as to accommodate variable redefinition. Lets consider the constraint **\*\*x:=p[i]*, x is a double pointer, p is single pointer, and i an integer. We assume **\*\*x* and **p* to be of assignment compatible types, for which the assignment operator is appropriately defined, either built-in or overloaded by user code. The resulting redefinition links and the single constraint relationship are shown in **Fig. 8**.

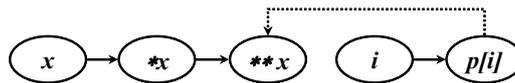

**Fig. 8.** Redefinition and constraint links due to constraint *\*\*x:=p[i]*.

For every declarative construct met, like the previous constraint, the code generation method will produce: (a) an initialization function for each distinct l-value; (b) a redefinition function for each redefining l-value; (c) the corresponding action function; and (d) a single initialization function for the entire declarative construct as a unit. The prototypes of such functions, with the implementation of the overall initialization function, are provided in **Fig. 9**; trivial name mangling is applied to exclude name conflicts. The *owner* parameter is null for globally defined declarative constructs, while it is supplied with the address of the owning object for declarative constructs defined within classes; for simplicity, we assume our example constraint is defined at global scope. Finally, the second *bool* parameter in all such functions plays a very important role: a *true* value leads to a typical installation behavior, while a *false* value causes the opposite behavior, i.e. cancellation.

```
init_sim_x(void* owner, bool b);            x        init_0(void* owner, bool b)
init_ptr_x(void* owner, bool b);           *x        {
init_ptr_ptr_x(void* owner, bool b);      **x          init_sim_x(owner, b);
init_sim_i (void* owner, bool b);           i          init_ptr_x(owner, b);
init_p_arr_i (void* owner, bool b);       p[i]         init_ptr_ptr_x(owner, b);
                                                       init_sim_i(owner, b);
redef_sim_x (void* owner, bool b);          x          init_p_arr_i(owner, b);
redef_ptr_x (void* owner, bool b);         *x        }
redef_sim_i (void* owner, bool b);          i
assign_0 (void* owner);
```

**Fig. 9.** The prototypes of the generated functions of our example (all of them *void*).

The initialization functions are produced have a twofold role: (a) manage the involvement of variables in declarative constructs; and (b) handle the 'privilege' of variables as redefining other variables. So, following our example, the initialization functions for **x* and *p[i]* handle their involvement in the constraint, while the initialization functions for *x*, **x* and *i* handle their role as redefining variables. The redefinition functions are produced to behave as follows: (i) firstly, they perform calls to the initialization functions of all their redefined variables; and (ii) only for those redefined variables that are also redefining other variables, they encompass explicit calls to their respective code-generated redefinition functions (this will ensure the recursive application of the redefinition behavior across the graph). For our example, these requirements are met by the code of **Fig. 10**.

```
void assign_0 (void* owner){ **x = p[i]; }
void init_sim_x (bool b)   { x.HandleRedefinition(redef_sim_x, b); }
void init_ptr_x (bool b)   { (*x).HandleRedefinition(redef_ptr_x, b); }
void init_ptr_ptr_x (bool b) {
       (**x).HandleConstraint(assign_0, b);
       (**x).HandleDependency(&(p[i]), b);
       if (b) assign_0();      ← Explicitly applying the constraint on initialization
}
void init_sim_i (bool b)   { i.HandleRedefinition(redef_sim_i, b); }
void init_p_arr_i (bool b) { }
void redef_sim_x (bool b)  { init_ptr_x(b); redef_ptr_x(b); }
void redef_ptr_x (bool b)  { init_ptr_ptr_x(b); }
void redef_sim_i (bool b)  { init_p_arr_i(b); }
```

**Fig. 10.** Code generation for the *init_* and *redef_* functions of our example; the *owner* parameter is omitted for clarity.

Following **Fig. 10**, the call graph due to updating *x* is provided in **Fig. 11**. The actions before change lead to uninstalling from **x* and ***x* the functions registered due to the example constraint, while the actions after change reinstall the same functions on the redefined **x* and ***x*.

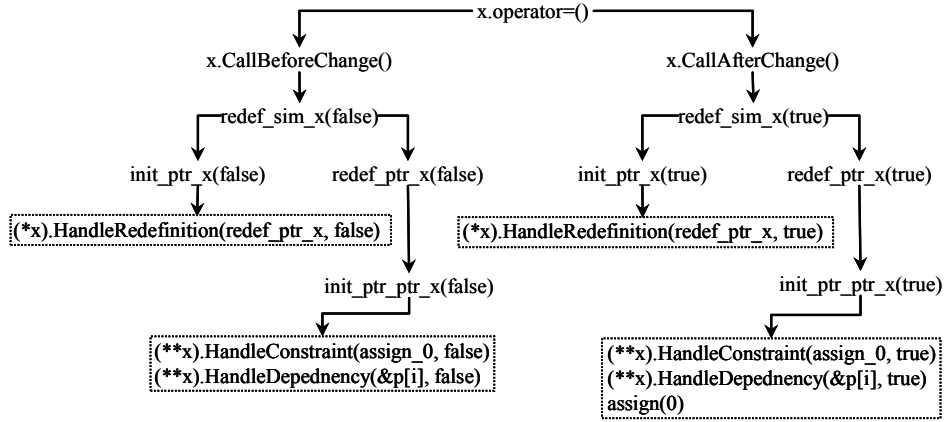

**Fig. 11.** The call graph resulting from updating *x*.

## 5. Code generator

The algorithm to generate the required *init_* and *redef_* functions is straightforward, relying on the redefinition information collected during parsing, both on the *redef* list of l-values redefining *l* as well as the list of l-values redefined by *l*. In this context, the code generator for redefinition functions, as well as for constraints, is provided within **Fig. 12**. The generator for pre-conditional statements and monitors is very similar to the *emit_constraint* function, requiring emitting the corresponding action function (like *emit_assign* for constraints) and calling the appropriate handler function of the *SmartSuperClass* API, like *HandleMonitor* or *HandlePrecondition*. The generator for redefining l-values works as already described: it emits the *init_* function installing the redefinition function and it emits the *redef_* function calling for initialization of all redefinition-dependent l-values, and for those also redefining other variables, calling as well their redefinition function.

*fg(l, id)* { **return** *id* + *mang(l)* + "`(void* owner, bool b){`"; }
*emit_assign(l, e)* {
   *s* = *newassignname()*
   **em** *s* + "`(void* owner) {`" + *l.str* + "`=`" + *e.str* + "`;}`"
   **return** *s*
}
*emit_redef(l)* {
   **em** *fg(l,*"`init_`"*)*+"`(`"+*l.str*+"`).HandleRedefinition(redef_`"+*mang(l)*+"`,b); }`"
   **em** *fg(l,* "`redef_`"*)*
   **for each** *x redefined by l* **do** {
      **em** "`init_`" + *mang(x)* + "`(owner, b);`"
      **if** *x redefines other variables* **then**
         **em** "`redef_`" + *mang(x)* + "`(owner, b);`"
   }
}
*emit_constraint(l, e)* {
   *a* = *emit_assign(l, e)*
   **em** *fg(l,* "`init_`"*)* + "`(`" + *l.str* + "`).HandleConstraint(`" + *a* + "`, b);`"
   **for each** *x in e.lvalues* **do**
      **em** "`(`"+*l.str*+"`).HandleDependency(&(`" + *x.str* + "`), b);`"
   **em** "`}`"
   *all* = { *l, e.lvalues* }
   **for each** *x in all* **do**
      **for each** *y in x.redef* **do**
         *emit_redef(y);*
   **em** "`}`"
}

**Fig. 12.** Code generator of redefinition functions and constraints.

Finally, we extend the rule constructing the *str* attribute of l-values to handle declarative constructs at class scope: at the rule *L.str*=**id**, if **id** is a member of the encompassing class *A*, we introduce the prefix "*(A\*) owner)->*". To ensure this type casting is safe, we have to call unit initialization (e.g. the *init_0* function of our main example) in every class constructor, passing *this* in place of the *owner* parameter.

## 6. Handling objects

At this point, our method covers the involvement in declarative constructs of scalar variables. Additionally, for arrays the solution is straightforward, by treating all array elements as explicitly involved. However, the method is not yet complete to handle for objects. For example, given the constraint *x:=a.f()*, *a* instance of class *A*, and *f* a public function of *A*, upon modification of *a* constraint resolution is performed, leading to the assignment *x=a.f()*. If *a* has multiple data members, the key issue is '*when **a** is considered as modified, and when are the actions due to its update performed*'. A naïve approach is to assume the object is modified with every member data update, as it is ignorant of the design semantics of class *A* actually defining the rules of correctness for instance *a*. More specifically, following Design by Contract, not every member modification of *a* leads to a state preserving the *A* class invariant, neither can it ensure that the precondition of *A::f* is always satisfied. We may have a sequence of internal state transitions for an object where each such intermediate state is 'illegal' externally, i.e. not fulfilling the invariant. In conclusion, we need to somehow tune our method to trigger notifications for object updates only when those imply externally correct states.

***Trapping updates preserving the class invariant*** Design by Contract is not built-in in C++, so we consider the following criteria for object correctness: we assume calls to public functions lead to correct object states, and we assume calls to private functions from friend classes are applied on correct object states. Following these rules, we generate code to suspend update notifications on entry to member functions, and to resume notifications on exit. We treat suspend inquiries in an accumulating fashion, i.e. *N* suspends must be followed by *N* resumptions to cause update notification. In **Fig. 13** we provide the specialized smart type for classes, from which all user-defined classes are derived, together with the code generation scheme for classes.

```
class SmartClassSuper : public SmartTypeSuper { protected:
    void Suspend(void) { ++n; }
    void Resume(void)
        { if (!--n) if (update) { update=false; CallAfterChange(); } }
    void SetUpdated(void)
        { if (!update) { update=true; CallBefforeChange(); } }
};
```

***class A { T₁ x; public: A(); T₂ f(); };***

```
class A : public virtual SmartClassSuper {   Code generation version of A class
    struct E { A* p; E(A* q): p(q){} ~E(){ p->Resume(); } };
    friend struct A::E;
    static void update (void* owner) { ((A*) owner)->SetUpdated(); }
    public:
    A() { x.HandleMonitor(this, update); ...user statements... }
    T₂ f() { Suspend(); E tmp(this); ...user statements... }
};
```

**Fig. 13.** Smart-type for classes and the code generation scheme to implement deferred object update notification, 'aligned' with transitions among design-correct states.

Following **Fig. 13**, every class encompasses a static monitor function, registered upon construction to all data members (members may of a class type too), informing for an update via *SetUpdated* the owner object. The first call to *SetUpdated* will also cause a call to *CallbeforeChange* (since the object itself may redefine other variables). We prefix the blocks of member functions with a *Suspend* call, ensuring that the last statement is always a *Resume* call via a temp object at function block scope, which upon destruction calls for resumption. The *Resume* function posts an update notification via *CallAfterChange* only if *SetUpdated* was previously called and the notifications are enabled.

***Handling contractual preconditions*** For expressions involving objects calling member functions, like *a.f()*, we have to ensure that such calls may only be committed, upon object modification, if their contractual precondition is also satisfied. From the three types of declarative constructs, monitors and pre-conditional statements do not need special treatment, as the statement on the RHS is typical imperative code that should be programmed to obey the contract. The issue arises on constraints which result in the automatic application of the constraint assignment. More specifically, given *x:=h(a.f(), b.g())*, even if *a* or *b* are modified, we have to ensure that the assignment is performed if the preconditions of *f* and *g* are both fulfilled. Since preconditions are not built-in in C++, we introduce constraints with preconditions, i.e. the constraint is applied only when its precondition is *true*, shifting responsibility to programmers. This extension affects slightly both the code generation, as constraint precondition functions must be generated, and the way constraint resolution is applied, since constrained objects are now updated if their constraining variables change and if their precondition is also fulfilled. In result, assuming the contractual preconditions *pre_f* and *pre_g* are defined for *f* and *g* respectively, the previous constraint should be programmed as:

$$x:=h(a.f(), b.g()) \text{ given } a.pre\_f() \text{ \&\& } b.pre\_g()$$

## 7. Summary and conclusions

We focus on a compact set of declarative programming elements, originated from User Interface programming languages, enabling a departure from the strict imperative programming of program control flow. From the implementation point of view, the mixing of such new elements with the C++ language becomes very demanding due to the nature of l-values in the C++ language. Our method proposes three highly generic forms of declarative elements, while it caters for l-value redefinitions and supports objects. The implementation complexity is split in two parts, the augmented type library responsible for update detection and propagation, and the code generation method relying on redefinition graphs to produce functions that handle variable redefinition during runtime.

Currently, we are investigating the way exception handling should be handled in declarative constructs, in particular, when an exception is raised due to the triggering condition of pre-conditional statements and the constraint assignment of constraints. One practical barrier of our method is the potential lack of backward compatibility. For instance, in our implementation, any application of memory copy operations among regions of scalar values, commonly met in C++ programs, is very hazardous, as it is actually applied on aggregate objects. We have tried to overcome this serious drawback with low-level techniques, like turning scalars to non-derived 'less smart' classes with data content and size equivalent to their native type, while dynamically allocating the 'smart' counterpart locatable via a hash table indexed with the address of the 'less smart' instance. However, from our initial tests, the performance overhead was disappointing and unacceptable; we are still working on this issue.

Overall, the reported work reflects our belief that we may pursue the evolution of general purpose languages with an in depth investigation and extrapolation of interesting elements met in special-purpose languages for relatively 'large' domains. Finally, although the reported implementation is C++ specific, the technique may also be applied to other languages like Java or C#.